\documentclass[reprint,
superscriptaddress,
 amsmath,amssymb,
 aps,
]{revtex4-1}

\usepackage{color,soul}
\usepackage{graphicx}
\usepackage{dcolumn}
\usepackage{bm}
\usepackage{inputenc}
\usepackage{csquotes}
\usepackage{subfigure}
\usepackage[colorlinks,allcolors=blue]{hyperref}
\usepackage{float}

\begin{document}


\title{Scattering Statistics in Nonlinear Wave Chaotic Systems}

\author{Min Zhou}
\affiliation{Department of Electrical and Computer Engineering, University of Maryland, College Park, Maryland 20742, USA}
\affiliation{Center for Nanophysics and Advanced Materials, University of Maryland, College Park, Maryland 20742, USA}

\author{Edward Ott}%
\affiliation{Department of Electrical and Computer Engineering, University of Maryland, College Park, Maryland 20742, USA}
\affiliation{Department of Physics, University of Maryland, College Park, Maryland 20742, USA}

\author{Thomas M. Antonsen, Jr}%
\affiliation{Department of Electrical and Computer Engineering, University of Maryland, College Park, Maryland 20742, USA}
\affiliation{Department of Physics, University of Maryland, College Park, Maryland 20742, USA}

\author{Steven M. Anlage}%
\affiliation{Department of Electrical and Computer Engineering, University of Maryland, College Park, Maryland 20742, USA}
\affiliation{Center for Nanophysics and Advanced Materials, University of Maryland, College Park, Maryland 20742, USA}
\affiliation{Department of Physics, University of Maryland, College Park, Maryland 20742, USA}
\date{\today}

\begin{abstract}
The Random Coupling Model (RCM) is a statistical approach for studying the scattering properties of linear wave chaotic systems in the semi-classical regime. Its success has been experimentally verified in various over-moded wave settings, including both microwave and acoustic systems. It is of great interest to extend its use to nonlinear systems. This paper studies the impact of a nonlinear port on the measured statistical electromagnetic properties of a ray-chaotic complex enclosure in the short wavelength limit. A Vector Network Analyzer is upgraded with a high power option which enables calibrated scattering (S) parameter measurements up to +43 dBm. By attaching a diode to the excitation antenna, amplitude-dependent S-parameters are observed. We have systematically studied how the key components in the RCM are affected by this nonlinear port, including the radiation impedance, short ray orbit corrections, and statistical properties. By applying the newly developed radiation efficiency extension to the RCM, we find that the diode admittance increases with excitation amplitude. This reduces the amount of power entering the cavity through the port, so that the diode effectively acts as a protection element.
\end{abstract}

\maketitle
\textbf{Many wave systems, ranging from the quantum mechanical wavefunctions of complex molecules to the reverberation of sound in a large concert hall, share a common description in terms of wave chaos.  Such systems have been shown to have universal statistical fluctuations governed by random matrix theory, but their properties are also “dressed” by non-universal features that are specific to each phenomenon.  Of particular interest are the scattering properties of wave chaotic systems that are open to the outside world through a finite number of scattering channels.  We have developed the Random Coupling Model (RCM) to provide a complete quantitative understanding of all such systems, and we have extended it in several ways to account for increasingly complicated features of such systems.  In this paper we further extend the RCM to understand the amplitude-dependent universal and non-universal properties of a wave chaotic system with a strong nonlinearity built into a coupling channel.  Using this extended RCM we are able to understand the amplitude-dependent experimental data on the scattering properties of a microwave cavity with a nonlinear diode attached to scattering port.  This is an important step in the ongoing effort to create the science of nonlinear wave chaos.}

\section{Introduction}
Concepts from the field of wave chaos have been shown to successfully predict the statistical properties of linear fields in enclosures with dimensions much larger than the wavelength. This includes the properties of closed systems, such as eigenvalues and eigenfunctions, as well as the scattering properties of open systems \cite{stockmann}. The Random Coupling Model (RCM) describes the scattering properties by incorporating both universal features described by Random Matrix Theory (RMT) \cite{stockmann,casati,boh,guhr_random-matrix_1998,beenakker_random-matrix_1997,alhassid_statistical_2000,haake_quantum_2000} and the system-specific features of particular system realizations \cite{SH-uni,XZheng,RCMreview}. The RCM is formulated in terms of the Wigner reaction matrix, directly analogous to the electromagnetic or acoustic impedance, rather than the scattering matrix \cite{beck_R-matrix_2003,mitchell_random_2010}. This allows the RCM to be expanded and appended in a simple additive or multiplicative manner, creating opportunities to describe increasingly complicated scattering scenarios. Examples of such complications include taking account of \enquote{short orbits} between the ports (or scattering channels) that survive the ensemble averaging process, \cite{James-shortray,yeh-SO1,yeh-SO2} and modifications of scattering statistics due to losses localized in the ports, rather than in the scattering system \cite{bisrat,bo,bisrat2}.

Nonlinearity in wave-chaotic systems has been studied in several aspects. For example, rouge waves can appear in linear wave chaotic scattering systems \cite{sea1, sea2}. However, such waves can also appear in a variety of physical contexts and are enhanced by nonlinear mechanisms \cite{rogue1, rogue2}. In acoustics, Time-Reversed Nonlinear Elastic Wave Spectroscopy (TR/NEWS) is based on the nonlinear time reversal properties of a wave chaotic system \cite{8}. TR/NEWS is proposed as a tool to detect micro-scale damage features (e.g., delaminations, micro-cracks or weak adhesive bonds) via their nonlinear acoustic signatures \cite{9,10}. Applying this idea to electromagnetic waves \cite{Fink}, the nonlinear electromagnetic time-reversal mirror shows promise for novel applications such as exclusive communication and wireless power transfer \cite{MF-PRL,MF-PRE,cangialosi_time_2016,roman_selective_2016}. Theoretical study of stationary scattering from quantum graphs has been generalized to the nonlinear domain, where the nonlinearity creates multi-stability and hysteresis \cite{NLgraph}. A wave-chaotic microwave cavity with a nonlinear circuit feedback loop demonstrated sub-wavelength position sensing for a perturber inside the cavity \cite{NL2D}. Nonlinearity is a key ingredient in various machine learning protocols, including neural networks \cite{lecun_deep_2015,satat_object_2017} and reservoir computing \cite{pathak_using_2017,pathak_model-free_2018}. Utilizing wave chaotic layers, along with nonlinearity, offers an attractive way to enable physical realizations of deep learning machines \cite{shen_deep_2017,rotter_light_2017,del_hougne_shaping_2017}.

Nonlinear effects in wave chaotic systems manifest as harmonic and sub-harmonic generation, driving amplitude dependent responses, etc. We have recently studied the statistics of harmonics generated in a wave chaotic system by adding an active frequency multiplier to the $\frac{1}{4}$-bowtie microwave billiard \cite{MZ-2f}, which is a vertically thin (less than a half-wavelength) microwave cavity whose horizontal shape resembles a quarter of a bowtie (Fig. \ref{fig1}). This is quite relevant to the work that investigates the electromagnetic field statistics created by nonlinear electronics inside a wave chaotic reverberation chamber, and it has a number of applications in the EMC (Electromagnetic Compatibility) community, such as electromagnetic immunity testing of digital electronics \cite{14,1NL}. Another approach to observe nonlinear effects is to create a scattering system with amplitude dependent response. To achieve this, we have introduced different sources of nonlinearity into the billiards, and in this paper we focus on a high frequency diode. Reaching the nonlinear regime usually requires high amplitude inputs, hence we have implemented a high power vector network analyzer (VNA) which is able to measure the scattering (S) parameters for signals up to $\sim$ +43 dBm ($\simeq$ 20 Watts). 

\section{RCM Overview}
The basic idea of the RCM is as follows. For an N-port ray chaotic system, the statistical properties of the $N\times N$ cavity impedance matrix $\bar{\bar{Z}}_{cav}$ are described by a universally fluctuating complex normalized impedance $\bar{\bar{\xi}}$ and the system specific properties $\bar{\bar{Z}}_{avg}$ through the following equation \cite{XZheng,RCMreview,James-shortray,yeh-SO1,yeh-SO2}
\begin{equation}
\bar{\bar{Z}}_{cav}=i\cdot Im\{\bar{\bar{Z}}_{avg}\}+[Re\{\bar{\bar{Z}}_{avg}\}]^{1/2}\cdot \bar{\bar{\xi }}\cdot [Re\{\bar{\bar{Z}}_{avg}\}]^{1/2},
\label{eq_RCM}
\end{equation}
where $\bar{\bar{Z}}_{avg}$ is the average impedance over an ensemble of cavity realizations. $\bar{\bar{Z}}_{avg}$ contains the non-universal features of the system, including the radiation impedance of the port (which fully characterizes the port-specific properties that determine the \enquote{prompt response} to an input excitation \cite{SH-uni,James-shortray,yeh-SO1,SH-S,SH-PRE06}), and short orbits that survive the ensemble average \cite{James-shortray,yeh-SO1,yeh-SO2}. The \enquote{radiation impedance} $\bar{\bar{Z}}_{rad}$ describes the impedance of the port when only outgoing waves are present. Practically speaking, it is measured in the case that the waves get into the cavity through the port but do not return. This can be realized for example by covering the boundary of the billiard with perfect microwave absorbers. A \enquote{short orbit} describes a ray trajectory that leaves the port and immediately returns to it, or another port, without ergodically visiting the chaotic system. It is the result of the port-to-port interaction that introduces deterministic field components which can remain fixed in the ensemble \cite{yeh-SO1}. Under the assumption that losses are uniform, the statistics of the universally fluctuating complex impedance $\bar{\bar{\xi}}$ is determined by a single parameter named the loss parameter $\alpha$ \cite{SH-uni,XZheng,SH-S,SH-IEEE}. For a two-dimensional electromagnetic system (i.e., a vertically thin cavity), it is given by $\alpha=k^2A/(4\pi Q)$ and can be interpreted as the ratio of the typical 3-dB bandwidth of the resonant modes to the mean mode spacing. Here $k=2\pi f/c$ is the wave number at frequency $f$, $A$ represents the area of the billiard, and $Q$ is the typical loaded quality factor of the enclosure. The loss parameter $\alpha$ can vary from 0 (isolated resonances) to infinity (many overlapping resonances). The universal statistical properties of a chaotic system with loss parameter $\alpha$ is given by RMT \cite{XZheng,RCMreview,Fyodorov,fyodorov_scattering_2005,li_random_2017}. Note that we can invert Eq. (\ref{eq_RCM}) and create an experimental approximate to the statistics of $\bar{\bar{\xi}}$, called $\bar{\bar{\xi}}_{exp}$ by gathering an ensemble of $\bar{\bar{Z}}_{cav}$, and constructing $\bar{\bar{Z}}_{avg}$. By fitting the statistics of $\bar{\bar{\xi}}_{exp}$ to theoretical predictions, the loss parameter of the system can be estimated.

The success of the principles underlying the RCM has been experimentally verified in linear wave chaotic systems including microwave systems \cite{mw1,mw2,mw3} and acoustic systems \cite{acR1,acR2,auregan_acoustic_2016, tanner_wave_2007,wright_acoustic_2010}, from 1D quantum graphs \cite{graph,ZFu,kottos}, 2D electromagnetic billiards \cite{yeh-SO1,yeh-SO2, SH-S,stockmann_quantum_1990}, and 3D cavities \cite{alt_wave_1997,Zach-3D,JGG,Li}. Based on its success and flexibility, it is of great interest to extend the RCM to other systems. One area of extension is to nonlinear systems. 

In this work, we show the results for measurements of the nonlinear scattering parameters in a diode-loaded $\frac{1}{4}$-bowtie quasi two-dimensional microwave cavity. The $\frac{1}{4}$-bowtie cavity is a ray-chaotic billiard that displays universal statistical properties predicted by RMT and RCM \cite{XZheng,RCMreview,Fyodorov,mw2,AliG,DHWu,SHCHung,XZhengPRE,SH-PRE06,SH-PRB06,JHYeh-PRE12}. In this case the diode acts as a nearly point-like nonlinearity in a wave chaotic system. Attaching a diode to the excitation port, we observed that the raw cavity statistics of the impedance change substantially with the excitation power. We extend the RCM to this situation and use it to analyze our experimental results. We find that when the radiation impedance becomes nonlinear, short orbits between the port and a nearby wall, and the raw impedance statistics are strongly modified.  We also find that many of these changes are due to the fact that the admittance of the diode changes with the excitation power. The nonlinear diode competes with the cavity admittance, substantially altering the response of the system. By implementing the lossy port model extension of the RCM \cite{bisrat,bo,bisrat2}, the results are well explained by the changing radiation efficiency of the diode-loaded port. As a result, The diode effectively acts like a protection element in this configuration.

\section{Experimental Setup}
In the small signal limit, our system can be approximated as linear. To observe a nonlinear response, the system must have some sort of nonlinear property, and a large excitation signal is required. In our earlier studies of wave chaotic systems with one port or multiple ports, we measured the scattering parameters and used these measurements to study the statistical properties of the system. Here we measure the high power S-parameters including a nonlinear element in the wave system, at power levels up to +43 dBm (supplemental material). 
\begin{figure}[H]
	\begin{center}
		\includegraphics[width=0.9\linewidth]{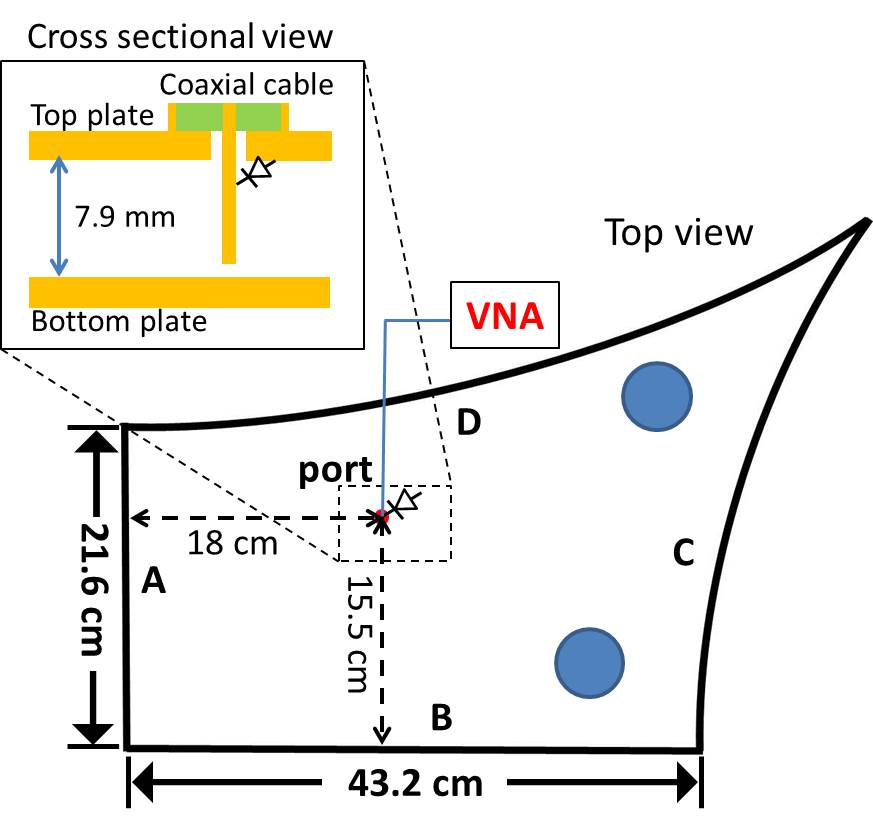} 
	\end{center}
	\caption{Top view of the experiment setup of the $\frac{1}{4}$-bowtie quasi-2D microwave billiard loaded with a diode attached to the single port. The diode (Infineon BAS7004) is connected between the center pin of the port and the top plate. The antenna pin is 7.6 mm long, 1.27 mm in diameter. The diode package has dimension $1.3\times 2.9\times1$ mm$^3$. The Vector Network Analyzer (Keysight N5242A PNA-X) measures the scattering parameter at excitation levels up to +43 dBm with microwave wavelengths from 3 to 7.5 cm. The two blue solid circles are metallic perturbers that can be moved around to create ensemble realizations. The inset shows a side-view cross section through the diode-loaded antenna.}
	\label{fig1}
\end{figure}

To induce strong nonlinearity, a diode (Infineon BAS7004 with two diodes in the package but only one is electrically connected) is soldered between the center pin and cavity ground, as shown in the inset to Fig. \ref{fig1}.  From the datasheet \cite{datasheet}, this diode has low transition capacitance, $C\sim1.5$ pF at 1 MHz, which decreases nonlinearly to $\sim$0.5 pF as the reverse voltage increases. Its differential resistance also changes nonlinearly as a function of the forward current. For typical forward currents $I_F=1\sim15$ mA, the resistance $R$ changes from 80 to 20 $\Omega$. A rough estimate for the time constant $\tau_{RC}=RC\sim100$ ps, which is close to the charge carrier life-time as given in the data sheet. Thus this diode can respond in the GHz frequency range and produces clear nonlinear responses, making it suitable for our microwave wave chaos experiments \cite{de_moraes_unified_2003,moraes_effects_2004}. In addition, the diode package is significantly smaller than the wavelengths used in this study (30-75 mm), rendering it approximately \enquote{point like}. The connection shown in Fig. \ref{fig1} has advantages in terms of stability and reproducibility, due to the fact that when the bowtie billiard is opened, the antenna and the top plate are attached together as one piece, and the bottom plate is a separate piece. This in turn allows for excellent reproducibility of $\bar{\bar{Z}}_{rad}$, $\bar{\bar{Z}}_{cav}$ and $\bar{\bar{Z}}_{avg}$ measurements.

\section{Results}
\subsection{Diode-Loaded Port Radiation Impedance}

\begin{figure}[H]
	\begin{center} 
		\subfigure{\label{fig2a}\includegraphics[width=0.6\linewidth]{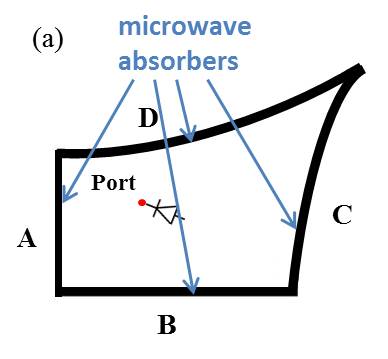}} 
		\subfigure{\label{fig2b}\includegraphics[width=1\linewidth]{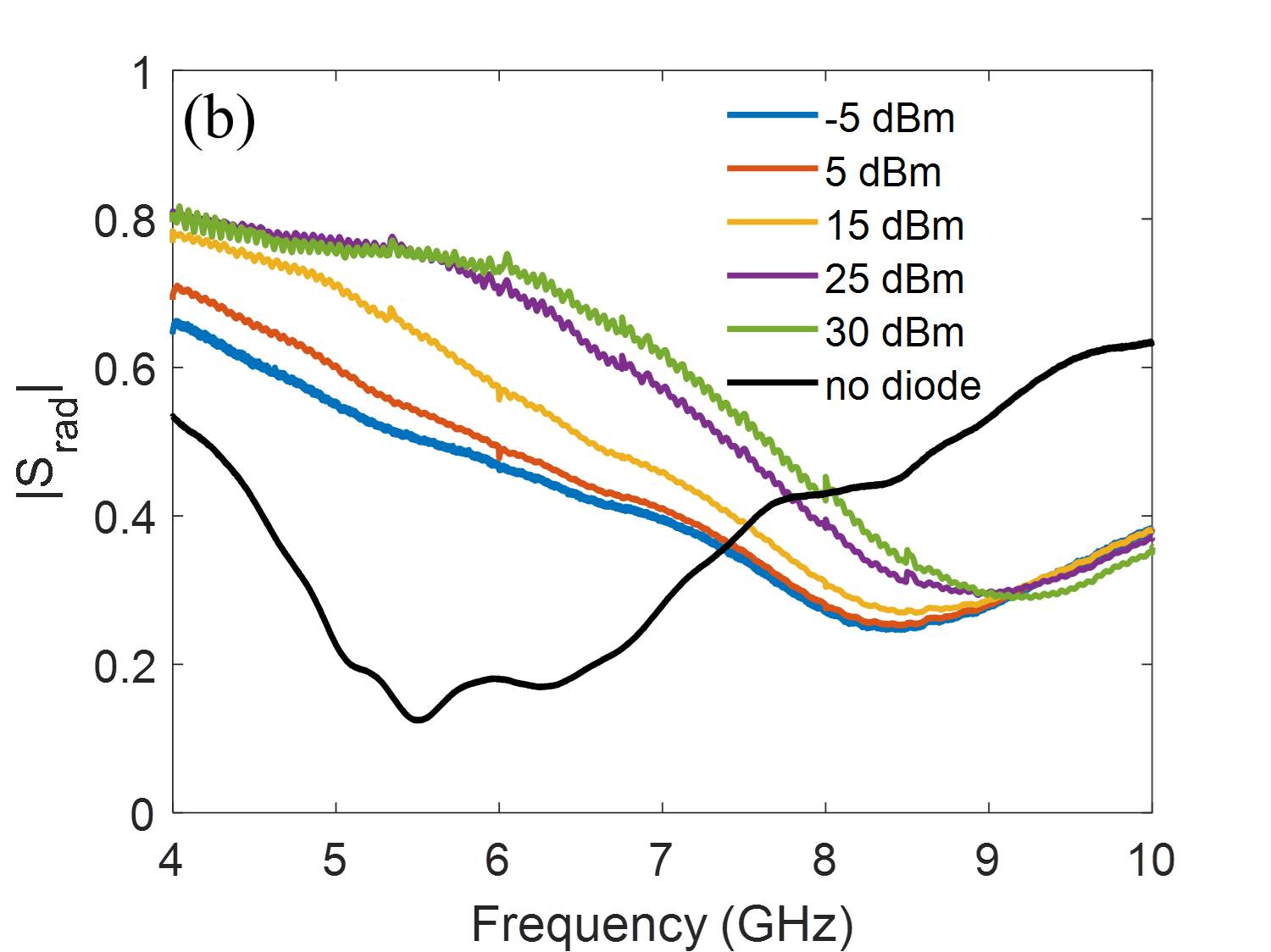}}
	\end{center}
	\caption{(a) Schematic illustration of the perimeter of bowtie covered with microwave absorbers to facilitate measurements of the radiation impedance with walls A, B, C and D labeled. (b) Measured $|S_{rad}|$ in the radiation case at different input powers, compared to the case of the antenna with no diode.}
	\label{fig2}
\end{figure}
\twocolumngrid
We first characterize the nonlinear port by measuring the radiation impedance. The radiation impedance characterizes the port properties alone. It is measured by creating an outward-only wave propagation condition in the experiment. This is achieved to good approximation by covering the perimeter of the bowtie billiard with microwave absorbers, as shown in Fig. \ref{fig2a}. Fig. \ref{fig2b} shows the measured radiation scattering parameter $S_{rad}$ at different input power levels, and includes the case of the antenna without the diode as well. Firstly, by adding the diode, the coupling of the port has been changed substantially. The optimal coupling near 6 GHz for the antenna has been moved to $8\sim9$ GHz when the diode is added. Comparing the nonlinear port radiation impedance at different power levels, there is a nonlinear change in the 4 $\sim$ 9 GHz range. The coupling is better at low power and the port generally reflects more as power increases.

\subsection{Short Orbits}
\begin{figure}[H]
	\begin{center}
		\subfigure{\label{fig3a}\includegraphics[width=0.49\linewidth]{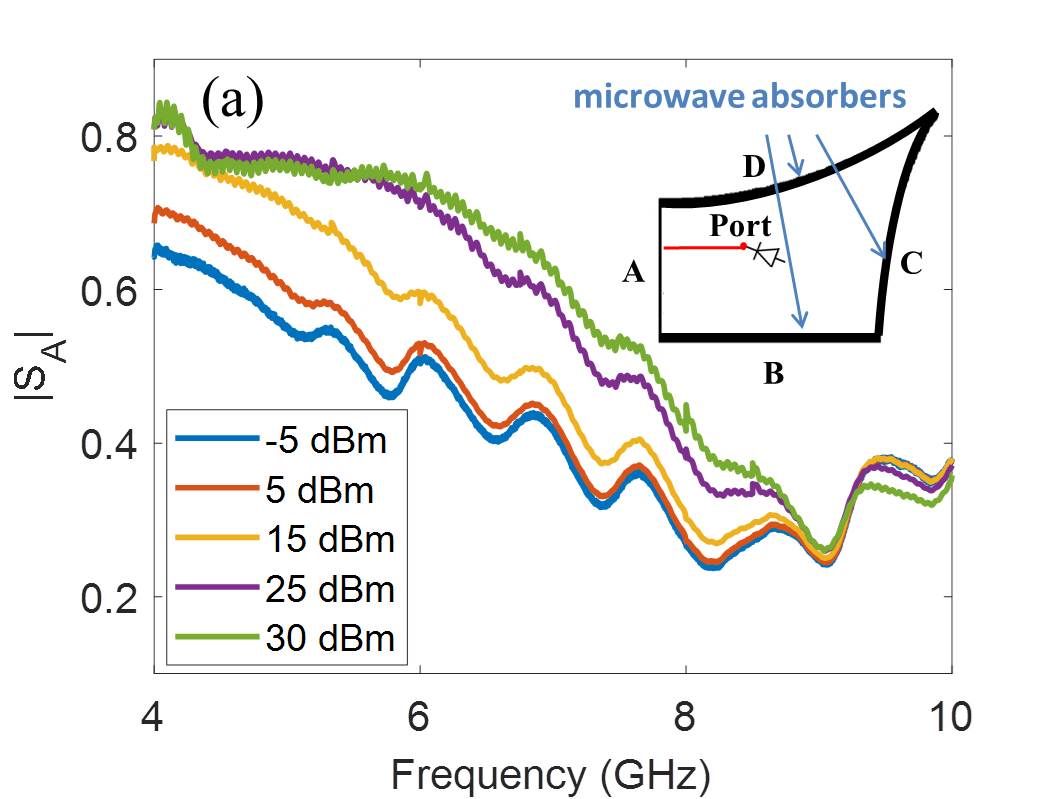}}
		\subfigure{\label{fig3b}\includegraphics[width=0.49\linewidth]{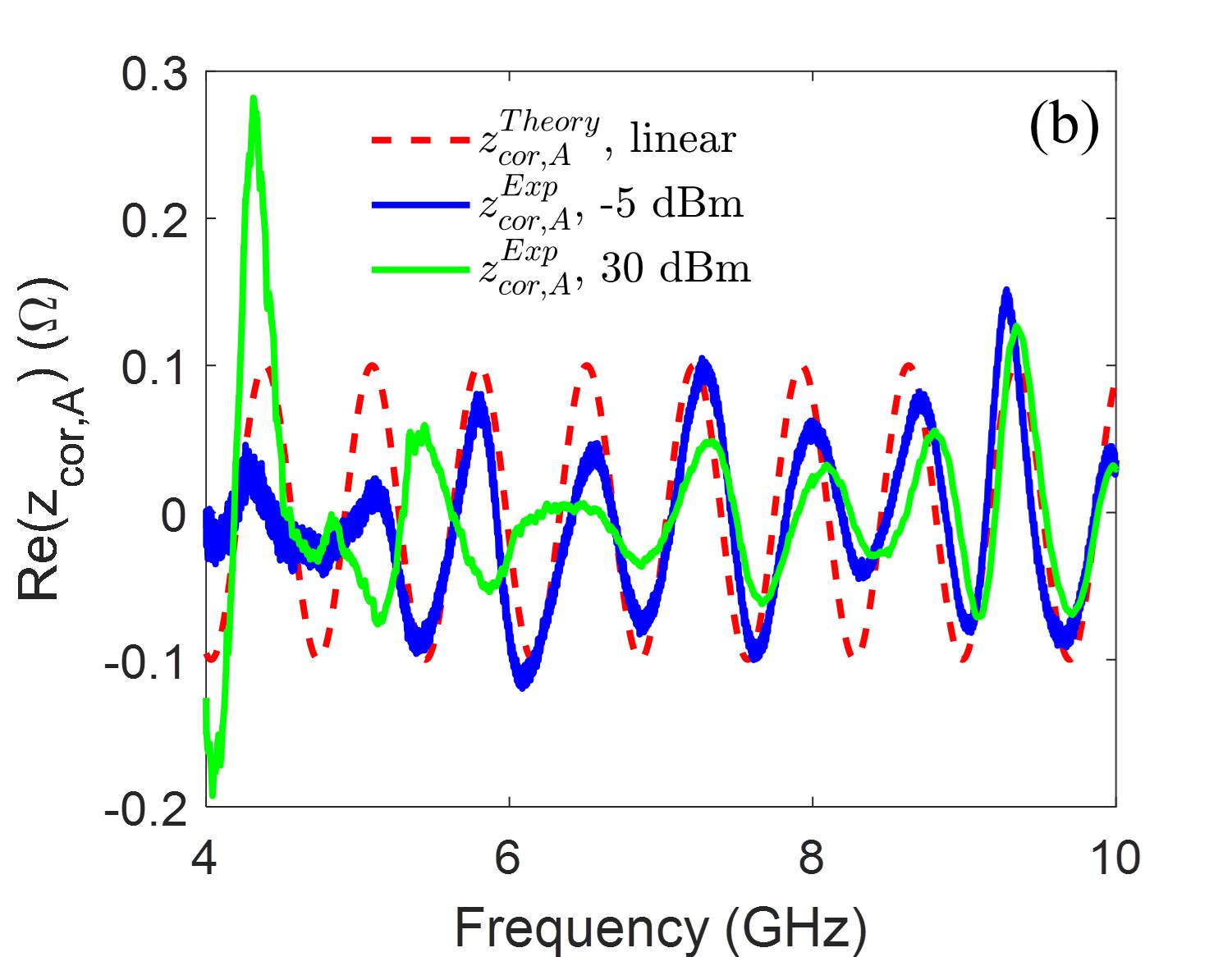}}
	\end{center}
	\caption{(a) $S_{A}$ measurement results as a function of frequency due to one short orbit between the port and wall "A" shown in the inset. The short orbit creates systematic variations in the S-parameters. The periodicity is related to the distance between the port and the exposed wall. The inset shows the experimental configuration for measuring the short orbit. The microwave absorbers are removed from wall A only, creating a single short orbit between the port and wall A. (b) Comparing short orbit corrections to impedance, $Re(z_{cor,A})$ from the experimental results for low power (+5 dBm, blue) and high power (+30 dBm, green) cases, with theoretical calculation (red), which assumes a linear response.}
	\label{fig3}
\end{figure}

Short orbits are another system specific feature that the RCM can incorporate for a more complete characterization of a complex scattering system. Short orbits between the ports can survive over an ensemble average because they appear in many or all of the realizations without modification. In refs. \cite{James-shortray,yeh-SO1,yeh-SO2}, the theory was developed and experimentally validated. The results for a single short orbit measurement are shown in Fig. \ref{fig3}. The inset of Fig. \ref{fig3a} is the experimental configuration for measuring a short orbit between the port and wall A of the billiard. The microwave absorbers of wall A are removed, so that there is a ray that goes into the billiard through the port and immediately reflects from wall A and goes back to the port. The direct measurements of the reflection S-parameter $S_{A}$ for this case are shown in Fig. \ref{fig3a}. We interprete the systematic wiggles in $|S_A|$ versus frequency (Fig. \ref{fig3b}) as arising from the short ray trajectory. The periodicity is a function of the distance between the port and wall A. The experimental short ray impedance correction is given by \cite{yeh-SO1,yeh-SO2}
\begin{equation}
\bar{\bar{z}}_{cor}^{Exp}=Re\{\bar{\bar{Z}}_{rad}\}^{-1/2}\cdot (\bar{\bar{Z}}_{wall}-\bar{\bar{Z}}_{rad})\cdot Re\{\bar{\bar{Z}}_{rad}\}^{-1/2},
\label{shortOrbit}
\end{equation}
where $\bar{\bar{Z}}_{wall}$ is the measured impedance of the billiard with specific wall(s) exposed, and $\bar{\bar{Z}}_{rad}$ is the radiation impedance as deduced from the data in Fig. \ref{fig2}. Theoretically, the short ray correction here is given by \cite{yeh-SO1,James-shortray,yeh-SO2}
\begin{equation}
z_{cor}^{Theory}=-exp[-(ik+\kappa)L-ikL_{port}-i\pi],
\label{SO2}
\end{equation}
where $z_{cor}^{Theory}$ is the correction to the impedance due to the short ray that goes into the billiard through port 1 and goes back through port 1, $\kappa$ is the effective attenuation parameter that takes account of propagation loss, $L$ is twice the distance between the nonlinear port and the exposed wall, which is 36 cm here, $L_{port}$ is the port-dependent constant length between the antenna and the input/output port. $L_{port}$ is typically 1-2 cm and is caused by the difference in location between the calibrated reference plane at the end of the VNA transmission line and the practical reference plane at the port antennas, due to the additional length in the SMA connector.
\onecolumngrid

\twocolumngrid

Fig \ref{fig3b} shows the comparison between the theory and experiment for the low power (-5 dBm) and high power (+30 dBm) cases. For the low power case, the periodicity of the short ray $z_{cor}^{Exp}$ agrees reasonably well with the theory, although the amplitude of the experimental curve varies with frequency. For the high power case, there is larger disagreement between the theory and the experiment, both in periodicity and amplitude. This indicates that this nonlinear port destroys the simple short orbit behavior, although at low power, it can still be approximated as a linear port. As will be shown below, the deviation at high power is because the diode admittance gets larger under high power excitation, such that it dominates over the billiard admittance. As a result, the short ray impedance correction is strongly modified. To some extent, this applies to the low power case as well. Another concern is that the short ray calculation assumes the port to be a point-like object. By attaching the diode, whose dimension is comparable to the antenna, this nonlinear port has a more complicated structure, rendering the port an extended object with no single \enquote{position} for the short-orbit calculation.

\subsection{Ensemble Realizations}
\begin{figure}[H]
	\begin{center}
		\subfigure{\label{fig4a}\includegraphics[width=0.49\linewidth]{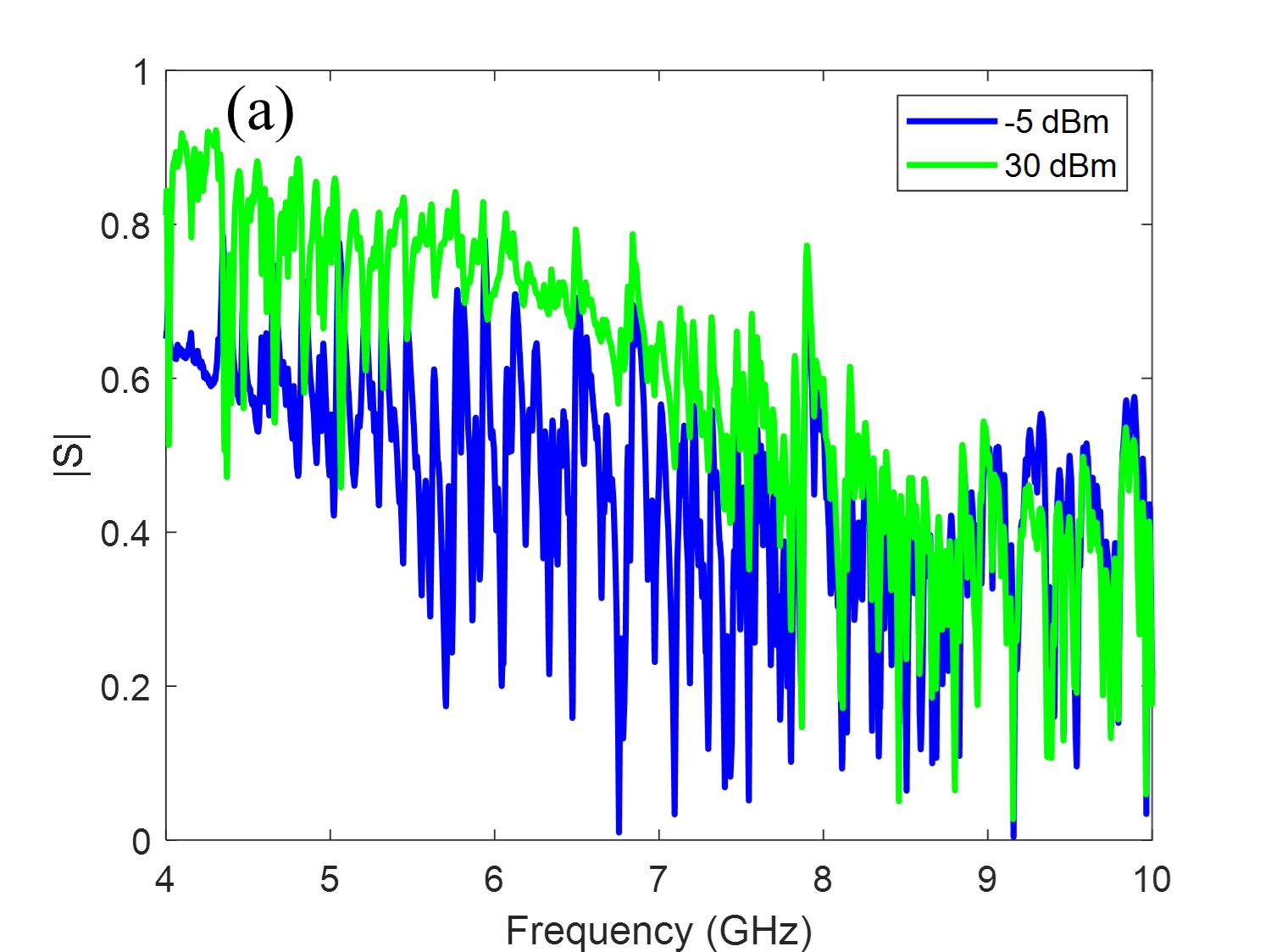}}
		\subfigure{\label{fig4b}\includegraphics[width=0.49\linewidth]{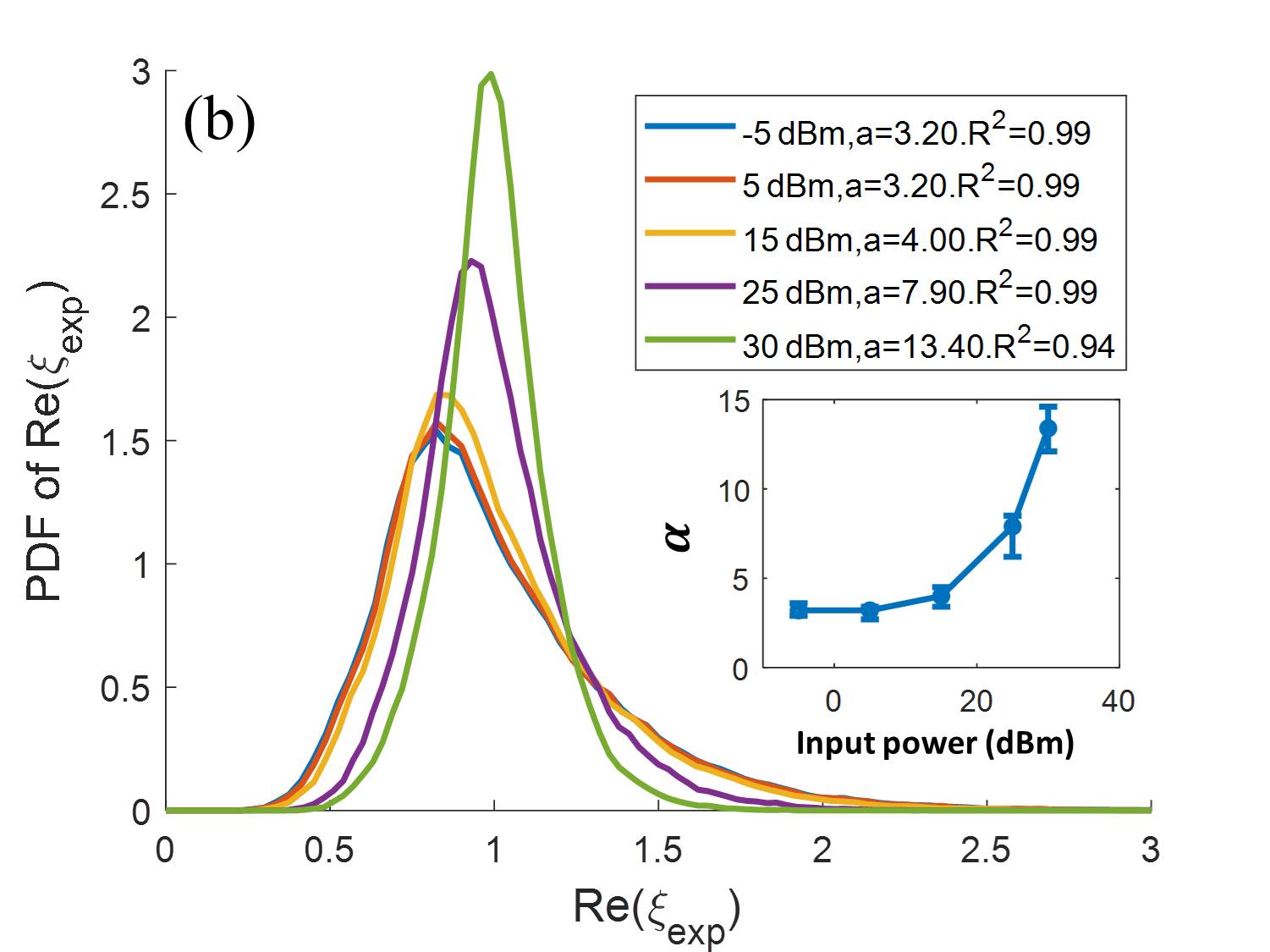}}
	\end{center}
	\caption{(a) Comparing the reflection S-parameter $|S|$ of a typical single realization of the 1/4-bow-tie cavity with diode-loaded port for low power (blue, -5 dBm) and high power (green +30 dBm). (b) Histogram of normalized $Re(\xi)$ obtained from ensemble data using traditional linear RCM for a 1 GHz window centered at 7 GHz. The resulting fitted loss parameter $\alpha$ increases with power as shown in the inset. $R^2$ values in the legend indicate the goodness of fit \cite{MZ-2f}.}
	\label{fig4}
\end{figure}

Having studied how the system specific properties change in the presence of the nonlinear port, we next analyze the statistics of the cavity itself. The microwave absorbers are removed from all of the walls of the billiard. The metallic perturbers shown in Fig. \ref{fig1} are moved around to create 120 distinct static realizations. Fig. \ref{fig4a} shows reflection vs. frequency results for a typical realization for low (blue) and high (green) input power. They have similar shapes as in the radiation case, but are \enquote{dressed} with many resonance fluctuations. The linear RCM approach applies well in the low power case, and we follow the RCM normalization process to determine an experimental approximate, $\xi_{exp}$ \cite{SH-uni,RCMreview,yeh-SO1}. Fig. \ref{fig4b} shows, PDFs of $Re(\xi_{exp})$ in the 6.5 to 7.5 GHz range for several different input powers. Clearly the PDF of the normalized impedance $Re(\xi_{exp})$ changes substantially with power, being more widely distributed in the low power case, indicating stronger fluctuations, a property which is associated with lower loss parameter $\alpha$. Note that the $Re(\xi_{exp})$ distribution is more concentrated near unity as power increases, consistent with a high loss (high $\alpha$) situation. If we naively fit this distribution function to the RCM using $\alpha$ as the sole fitting parameter, the fitted loss parameter $\alpha$ increases with power, as shown in the inset. The raw statistics of the system change substantially with power because of the presence of the nonlinear port. However it should be noted that these $\xi_{exp}$ PDFs show substantial deviations from RMT predictions (note the low fit $R^2$ values at high power), making it clear that naive application of the RCM breaks down in the nonlinear regime.
\onecolumngrid

\twocolumngrid

\subsection{Radiation Efficiency of the nonlinear port (High Loss System)}
\begin{figure}[H]
	\begin{center}
		\subfigure{\label{fig5a}\includegraphics[width=0.49\linewidth]{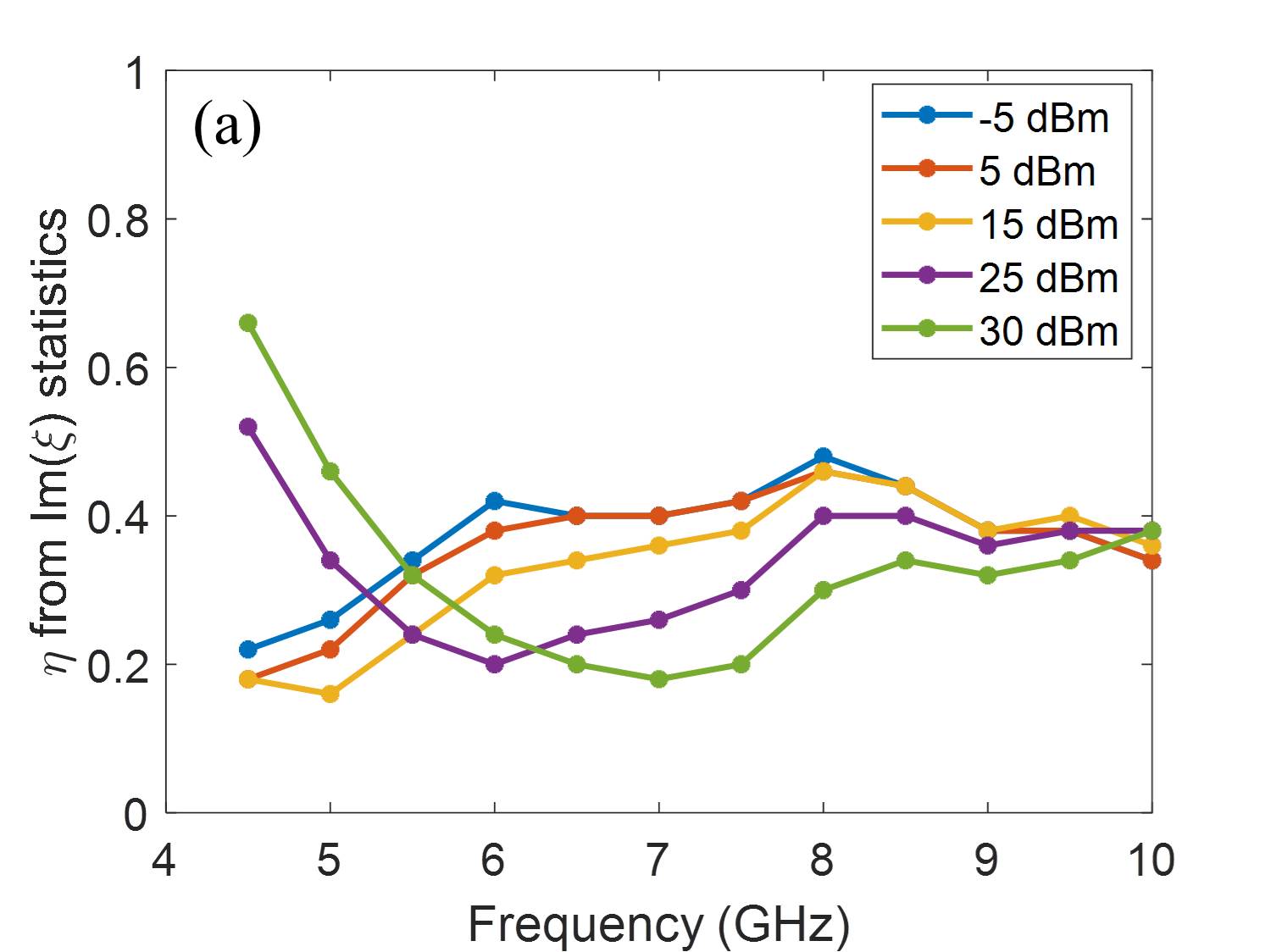}} 
		\subfigure{\label{fig5b}\includegraphics[width=0.49\linewidth]{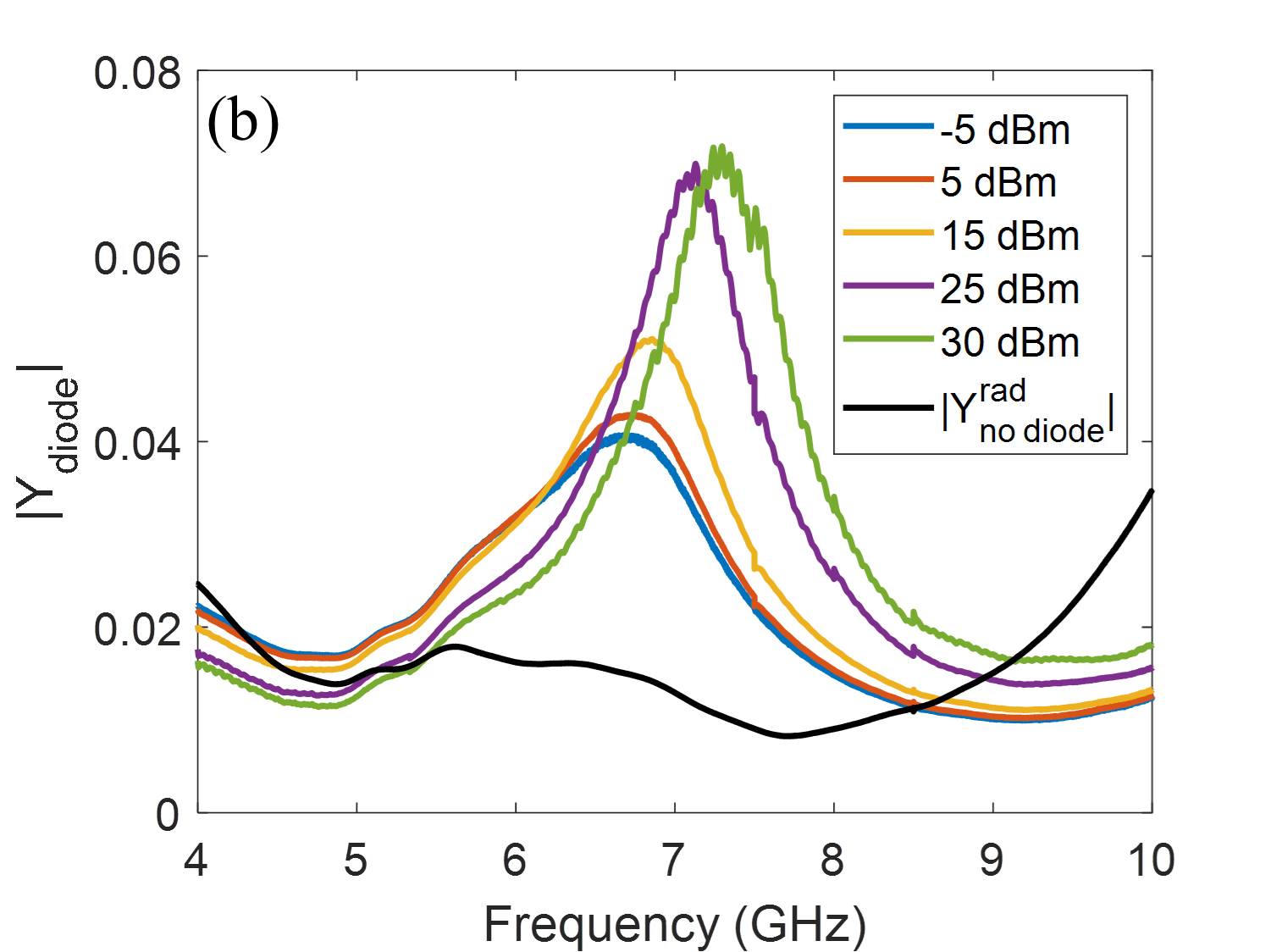}}
	\end{center}
	\caption{(a) Fitted radiation efficiency $\eta$ (from $Im(\xi)$ statistics) vs. frequency and power. Each fit was done with data from 120 realizations and a window of 1 GHz. (b) Plot of diode admittance magnitude vs. frequency at various rf powers, as well as the radiation admittance of the linear port $|Y_{no\ diode}^{rad}|$}	
	\label{fig5}
\end{figure}

In the RCM treatment presented above, we expect the loss parameter of the system to be independent of the excitation power as long as the properties of the cavity remain unchanged. The nonlinear property in this case is only associated with the port. The RCM described in Eq. (\ref{eq_RCM}) is derived assuming a lossless linear port. But this is no longer the case in this experiment. As we can see from Fig. \ref{fig4a} particularly in the vicinity of 6 GHz, the fluctuations are suppressed in the high power case, indicating that excitations of the cavity modes is suppressed. In this case, the port must be considered as a lossy port. Ref. \cite{bisrat, bisrat2} have derived a generalization of the RCM to account for the loss of the port. A radiation efficiency $\eta$ is introduced to quantify the ratio of the power radiated by the port to the input power to the antenna, $\eta=P_{rad}/P_{in}$, ($\eta$ is real and $0 \leqslant \eta \leqslant 1$). In a high loss system (i.e. $\alpha \gg 1$), it can be shown that the impedance of a lossy antenna inside a complex enclosure can be approximated as
\begin{equation}
\bar{\bar{Z}}_{in}=\bar{\bar{Z}}_{ant}+\eta \cdot Re\{\bar{\bar{Z}}_{ant}\} \cdot \delta\bar{\bar{\xi}},
\label{lossyPort}
\end{equation}
where $\eta$ is the radiation efficiency of the antenna, $\delta\bar{\bar{\xi}}= \bar{\bar{\xi}}-\bar{\bar{I}}$, $\bar{\bar{I}}$ is the identity matrix with diagonal elements $1+i0$, and $\bar{\bar{Z}}_{ant}$ is the input impedance of the lossy antenna radiating in free space. Ref. \cite{bo} has successfully applied this model to a scaled cavity, where the radiation efficiency accounts for the loss in free-space propagation suffered through a remote injection path. In our case $\bar{\bar{Z}}_{in}$ can be considered as $\bar{\bar{Z}}_{cav}$ and $\bar{\bar{Z}}_{ant}$ can be considered as $\bar{\bar{Z}}_{avg}$, therefore Eq. (\ref{lossyPort}) can be modified as
\begin{equation}
\bar{\bar{Z}}_{cav}=i\cdot Im\{\bar{\bar{Z}}_{avg}\}+(\bar{\bar{I}}+\eta\cdot  \delta\bar{\bar{\xi}})\cdot Re\{\bar{\bar{Z}}_{avg}\},
\label{lossyRCM}
\end{equation}
which is valid in the limit $\alpha \gg 1$. To determine $\eta$ for the nonlinear port, we first measure $\bar{\bar{\xi}}_{cav}$ of the billiard when there is no diode attached to the antenna. In that case $\bar{\bar{\xi}}_{cav}$ describes the properties of the billiard alone, (because all system-specific properties have been removed), and as such it is a linear system. We use the linear RCM approach, creating 120 realizations with the two perturbers, then applying Eq. (\ref{eq_RCM}) to extract $\bar{\bar{\xi}}_{cav}$ and fit to RCM to find the corresponding loss parameter $\alpha$ \cite{yeh-SO1, RCMreview, SH-uni}. Additionally, to make the bowtie billiard a high loss system, the perimeter of the billiard is partly (but uniformly) covered with microwave absorbers, tuning the loss parameter to be $\alpha=3$ (4.5 GHz) to $\alpha=7$ (9.5 GHz). Then using the value of $\alpha$ which characterizes loss in the cavity and $\bar{\bar{\xi}}_{cav}$, we go back to the diode-loaded nonlinear port case, utilize Eq. (\ref{lossyRCM}) and vary $\eta$ so that the statistics of $\bar{\bar{\xi}}(\eta)$ agrees with $\bar{\bar{\xi}}_{cav}$. Fig. \ref{fig5a} shows the fitted radiation efficiency from $Im(\xi)$ statistics, (the $Re(\xi)$ statistics give similar results).

Fig. \ref{fig5a} shows that the radiation efficiency is strongly power-dependent in the frequency range $4 \sim 10$ GHz. Between 6 and 9 GHz, the radiation efficiency decreases with increasing power, meaning the port is getting more lossy as the excitation power increases. There is a cross-over regime at low frequency 4 $\sim$ 6 GHz, where the radiation efficiency increases at high powers. And although it is not shown in the plot, at 10 GHz and above, the radiation efficiency is almost independent of power. 
\onecolumngrid

\twocolumngrid

To explain the results, we are interested in characterizing the diode admittance under different excitation powers. The diode is connected between the center pin and the ground of the port. The billiard radiation admittance can also be considered as being connected between the center pin and ground. Therefore a simple model is constructed by considering the diode and the billiard to be connected in parallel. By measuring the port radiation admittance with ($Y_{with\ diode}^{rad}$) and without the diode ($Y_{no\ diode}^{rad}$), the diode admittance can be approximated as $Y_{diode} \simeq Y_{with\ diode}^{rad}-Y_{no\ diode}^{rad}$. 

Fig. \ref{fig5b} shows the experimental results of $|Y_{diode}|$ compared with $|Y_{no\ diode}^{rad}|$. Between 4 $\sim$ 6 GHz, $|Y_{diode}|$ has similar values to $|Y_{no\ diode}^{rad}|$ and changes little with power. Thus the diode admittance competes roughly equal with the billiard admittance. A small change in admittance value may result in a big change in fitted radiation efficiency. However, between 6 and 9 GHz, the diode admittance dominates the billiard admittance and generally increases with power. In this frequency range, because the diode admittance is much larger than the billiard admittance and increases with power, the radiation efficiency of the port decreases, consistent with the results in Fig. \ref{fig5a}.

To better understand the nonlinear port, we built a circuit model for simulation in the finite difference time-domain code called CST(Computer Simulation Technology). The SPICE models of the diode as well as the package are provided by the manufacturer. In addition to the SPICE models, the dielectric properties of the package near the port also affect the radiation impedance, and this was added to the model. The internal capacitances of the package SPICE model were altered because the diode is being used beyond its design frequency range. The simulated amplitude dependent radiation S-parameters show relatively good agreement with the experimental results (see supplemental material). In addition the resultant radiation efficiency that is directly calculated as $P_{rad}/P_{in}$ is in general agreement with the experimental results. Based on the circuit model, the nonlinearity arises from the diode, which is approximated with an exponential I-V diode function \cite{wu_electromagnetic_2000}. The diode nonlinearity is shorted by parasitic capacitance in the package SPICE model at high frequencies, thus the port model does not have power dependence at 10 GHz and above, consistent with the data.

\section{Discussion and Future Work}
We have shown that by attaching a diode to the center pin and ground of an antenna, the port shows dramatic nonlinear behavior. By measuring the radiation impedance, which characterizes the port properties, we find the impedance changes considerably with the excitation power. The nonlinearity mainly occurs below 10 GHz, which is consistent with the diode time constant ($\tau\sim100$ ps, $1/\tau\sim10$ GHz). \enquote{Short orbits} are another system specific property. At low power, the short orbit behavior can still be approximately explained by theory, treating the port as a linear point source. As power increases, the short orbit correction deviates from the theoretical prediction. As the radiation efficiency fitting results show, this is because at high power the diode admittance dominates over the billiard admittance. 

For the statistical results, if we blindly apply the RCM to the billiard with diode-loaded port, one finds that the loss parameter increases with input power. The RCM posits that the loss parameter determines the universal properties of the chaotic system. In our case the billiard properties should not change with power because the nonlinearity is only associated with the port. We applied the newly developed radiation efficiency model to the port, using the radiation efficiency $\eta$ to quantify the proportion of power from the source radiated into the billiard. At high power in the nonlinear region, the diode consumes most of the power, causing the radiation efficiency to decrease. The diode thus prevents high power signals from getting into the billiard \cite{garver_diode_1976}.

There are several interesting questions for further study of this system. First, the billiard had to be intentionally modified into a high loss system in order to use the radiation efficiency model. The statistics of low-loss nonlinear systems cannot be addressed at this time. In addition, another behavior we have observed is the loss of reciprocity in a two-port version of this system, where one port is a nonlinear port and the other is linear. We observed that $S_{12}\neq S_{21}$ when large amplitude signals are applied. We note there is no general reciprocity theorem that holds for nonlinear systems \cite{reci}. This behavior can be understood by considering that with equal powers injected in both ports, the diode will be driven to nonlinearity when power is injected in the port hosting the diode, and to a lesser extent when the linear port is excited. 

Besides the approach we used in this paper to analyze the nonlinear system, we mention for completeness that high power S-parameters are sometimes called hot S-parameters \cite{dancila_solid-state_2014}, and the nonlinear effects can be fully characterized by the so-called X-parameters measured by the nonlinear VNA \cite{Roblin_nlVNA_2011,Root_xpara_2013}. However we believe that the present treatment is best suited for understanding the statistical properties of nonlinear wave chaotic systems in the semi-classical regime.

\section{conclusions}
To conclude, a diode based nonlinear port alters the radiation impedance, short orbits, and raw impedance statistics of a wave chaotic systems from those observed in linear systems. By using the Random Coupling Model with incorporation of the diode nonlinear properties and a lossy port model, these nonlinear phenomena are well explained, and verified by the simulation. The nonlinear property of the port can be applied to protect delicate circuits from high power electromagnetic microwave interference.

\section{SUPPLEMENTARY MATERIAL}
 The supplemental material contains more information on the high-power vector network analyzer
(VNA), and simulation results of the diode-loaded port in computer simulation technology (CST).

\section{ACKNOWLEDGMENTS}
This work is supported by the AFOSR under COE Grant No. FA9550-15-1-0171, the ONR under Grant No. N000141512134, the COST Action IC1407 \lq{ACCREDIT}\rq{} supported by COST (the European Cooperation in Science and Technology), and the Maryland Center for Nanophysics and Advanced Materials (CNAM).

\end{document}